# FLOW Mapping: Planning and Managing Communication in Distributed Teams

Kai Stapel, Eric Knauss, and Kurt Schneider
Software Engineering Group
Leibniz Universität Hannover
Hannover, Germany
{eric.knauss,kurt.schneider,kai.stapel}
@inf.uni-hannover.de

Nico Zazworka
Fraunhofer Center for Experimental Software Engineering, Maryland
College Park, USA
nzazworka@fc-md.umd.edu

*Abstract*—Distributed software development is more difficult than co-located software development. One of the main reasons is that communication is more difficult in distributed settings. Defined processes and artifacts help, but cannot cover all information needs. Not communicating important project information, decisions and rationales can result in duplicate or extra work, delays or even project failure. Planning and managing a distributed project from an information flow perspective helps to facilitate available communication channels right from the start – beyond the documents and artifacts which are defined for a given development process. In this paper we propose FLOW Mapping, a systematic approach for planning and managing information flows in distributed projects. We demonstrate the feasibility of our approach with a case study in a distributed agile class room project. FLOW Mapping is sufficient to plan communication and to measure conformance to the communication strategy. We also discuss cost and impact of our approach.

## I. INTRODUCTION

Global Software Engineering (GSE) research showed that many distributed projects face communication problems [1-5]. In a distributed setting collaboration is restricted by a limited set of communication media [3, 4, 6]. This may lead to communication breakdowns, misunderstandings, and finally to project delay or failure. Solutions have been proposed and evaluated to overcome communication problems in distributed development [4-7]. Although quite a few contributions found informal communication to be a main driver of global software projects [1, 2, 6, 8-10], not many approaches explicitly take informal, verbal, or ad-hoc communication into account. Informal communication is of importance especially at project start, when distributed teams still grow together and awareness and mutual understanding has to be developed [1, 6, 8, 11]. Incorporating informal communication when planning and managing a distributed project can thus help to get projects up and running faster and improve overall communication [11].

Informal communication does not happen as naturally as in co-located projects, therefore we believe a systematic method approach is needed for (1) planning and (2) managing communication in distributed teams. In our opinion such an approach should help to:

a) Capture information needs. Anticipate the different types of formal and informal communication activities of a project.
b) Specify desirable information flows. Choose adequate communication media for each communication activity.
c) Create a map of important information flows. Document the communication strategy for a project.
d) Apply the map of information flows. Implement the communication strategy.
e) Monitor information flows. Measure compliance to the communication strategy.

In the FLOW research project [12] we develop techniques to improve software development by specifically taking informal information flows into account. In this paper, we present the FLOW Mapping technique (Section III) to support planning and management of information flows in distributed projects. The central concept of FLOW Mapping is the FLOW Map (Section III.A), which helps to document important aspects of information flows that are missing in other notations (e.g. type of flows, including informal). Thus, the FLOW Map satisfies the need for documentation of a communication strategy (c) and is the primary document for training (d). As important as the FLOW Map is the method of creating it (a) and (b). In this paper we present the steps needed and the factors to take into account when creating a FLOW Map (Section III.B). The last part of FLOW Mapping presented here is an approach to measure conformance to the communication strategy. This approach helps to define how developers should implement the communication strategy (d) and to measure how well they perform it (e). Being able to measure how good a strategy performs is part of managing communication (Section III.C). Actions to take in case of non-conformance are out of scope of this paper and left to future research.

To test the feasibility of FLOW Mapping we performed it in a distributed agile classroom software project. We report on our experiences and measurement results in using FLOW Mapping in Section IV. During the case study we investigated how helpful FLOW Mapping and the resulting FLOW Map was for planning and managing communication in distributed teams. Compliance levels to the communication strategy varied between 79 % and 88 % for the presented conformance measures. Thus, we conclude that the communication strategy as planned was accepted by the developers. Before we present our concepts we start with a discussion of related work.

## II. RELATED WORK

We consider 3 areas of research to be related to the work presented here: 1) planning and managing communication in GSE projects, 2) factors that influence media choice in distributed settings, and 3) notations to visualize GSE projects.

*1) Managing Communication in GSE:* In a survey of 29 members of software development teams from 6 different countries Mohapatra et al. [13] identified communication mechanisms to be the most influential construct affecting coordination effectiveness in GSE. In a systematic literature review da Silva et al. [5] identified communication to be the most frequently dealt with challenge in managing distributed software development projects. The top five challenges revealed by the review are communication, cultural differences, coordination, time zone differences, and trust. In accordance to communication being a main challenge in GSE, the review found that the most solutions in the form of best practices and tools deal with communication problems as well. These results show that focusing on managing communication in GSE projects is promising. Our solution goes beyond best practices and tools by providing a systematic method for planning and management of GSE projects.

*2) Media Choice:* From analysis of 57 interviews from eight GSE projects Niinimäki et al. [14] identified factors affecting text-based versus audio-based communication media choice. The main factors are the communicator's role, level of language skills, communication context, task properties, media availability, and communication performance and satisfaction. For example, one result of their analysis is that technical personnel tend to prefer text-based over audio communication. In terms of level of language skill they discovered that self-conception of poor language skills also leads to the preference of text-based communication. This is just one out of many studies that identified factors that influence the choice of a communication medium. Inadequate media choice may be the result when developers are left on their own choosing a medium. Hence, we incorporate the choice of adequate communication media when planning a project, at least for the most important communication activities, in our approach.

*3) Visual Notations:* Laurent et al. [15] propose a visual modeling notation for planning distributed requirements engineering projects. Similarly to our FLOW notation (see Section III Concepts) this notation has elements for visualization of development sites, project participants, documents and communication flow. It also is designed to be intuitive to use and thus offers a limited set of symbols. Among the differences to the FLOW notation are its specialization on requirements engineering, more visually complex symbols that demand tool support, and the ability to depict communication media, type of document and different roles of participants. Laurent et al. report that they used the models to plan and assess communication flows as well as to use them as a basis to search for communication related problems.

## III. CONCEPTS

In this section we present our approach for setting up and measuring communication in distributed software development projects. We start with an introduction of the (A) core concepts of FLOW and the GSE specific extensions FLOW Maps and FLOW Mapping. After that we present the activities for the two phases of (B) planning and (C) managing communication.

### A. FLOW Mapping and FLOW Maps

FLOW Mapping is a technique from the FLOW Method [12] to *plan* and *steer* communication in distributed development projects. To achieve these goals the technique is centered around the visualization of a FLOW Map. A FLOW Map is a special FLOW model [16] (i.e. visualization of project participants, documents, and information flows) extended by features to improve awareness in distributed teams. Hence, a FLOW Map comprises features of FLOW and is extended by GSE specifics as first proposed in [11]. For the case study presented in Section IV we additionally added agile features. The following lists give a brief overview of the basic, GSE, and agile features of a FLOW Map as needed to understand the rest of this paper. For more exhaustive descriptions see [11, 12, 16].

The basic FLOW features are:
- Metaphor of the state of information to distinguish between information that is *solid*, i.e. (1) long term accessible, (2) repeatable accessible, and (3) comprehensible by third parties, or *fluid*, i.e. information that violates one of the solid criteria. Typical representatives for solid information are formal documents or formal e-mails. Fluid information on the other hand is knowledge in people's minds, or media that strongly depends on people's knowledge for correct interpretation like informal notes, informal e-mails, or the content of conversations e.g. via telephone.
- A simple graphical FLOW notation exists to visualize FLOW models [16].
- The notation has representations for solid information storages and flows and for fluid information storages and flows respectively. See Figure 2.

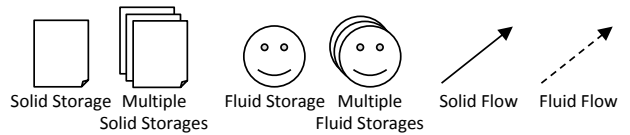

Solid Storage   Multiple Solid Storages   Fluid Storage   Multiple Fluid Storages   Solid Flow   Fluid Flow

Figure 2. Basic FLOW notation

FLOW Map extensions for GSE are:
- Assignment of storages to locations, e.g. development sites. The location of a fluid storage represents the physical location of the person possessing the information. The location of a solid storage on the other hand represents the site that is responsible for the content of the solid information storage, i.e. *not* its physical location.
- Line width emphasizes strength of information flow.
- Undirected connections represent information flows in both directions.
- Icons indicate communication media used for information flowing across sites.
- Additional meta-information about each storage, the so called yellow pages: contact information, picture, local time, status information, role in project, skills relevant for project, current task or work item, etc.

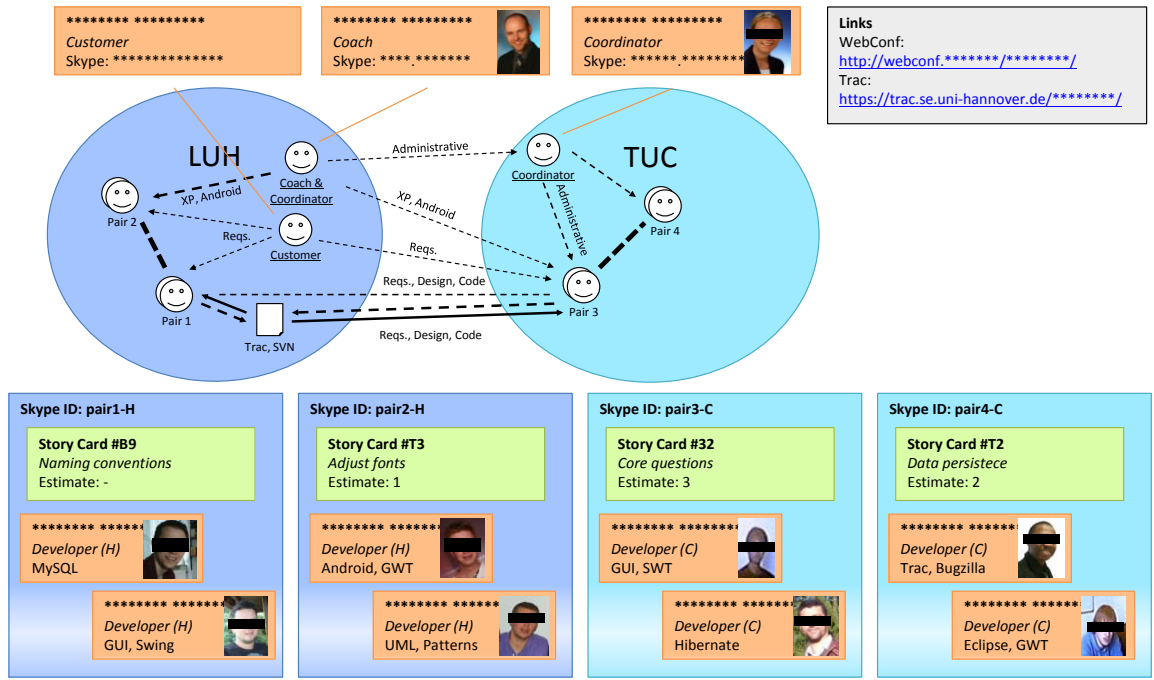

Figure 1. Target FLOW Map for project of the case study

FLOW Map extensions for agile development are:
- A double fluid storage symbol for pair programmers.
- Extended yellow pages information for pairs including their current work item (e.g. User Stories).

Figure 1 shows an example of a FLOW Map that was used in our case study. The FLOW model in the center depicts the participating developers and other stakeholders at their different locations along with desired information flows. At the top and bottom yellow pages information are provided for names, pictures, roles, Skype-IDs, programming skills, and current work item (i.e. User Stories).

A FLOW Map can be used as a cognitive tool for *planning* as well as for *managing* communication in a distributed project. Planning and management of a project based on the FLOW Map consists of a set of activities. We call the execution of these activities FLOW Mapping. In the following two sections we provide more details on the activities for planning (B) and for managing (C) communication using the FLOW Map.

### B. Planning Communication

In this section we describe how communication can be planned by creating a FLOW Map:

*1) Establish team:* Before starting a distributed project the coordinator of the project needs to establish the team. She needs to decide what sites/locations are participating and for each site what developers will be part of the distributed team. Constraints like the common spoken language or minimum development skills/experience need to be set as well. Finally, contact information of all participants need to be collected.

*2) Create Communication Strategy:* We define a communication strategy to be a set of scheduled or event-driven communication activities and a mapping between these activities and communication media to be used during their execution. A communication activity is an activity with specific information exchange goals. Example communication activities of an eXtreme Programming (XP) project are the planning game or daily stand-up meetings. The goal of a planning game is to get prioritized requirements from the customer. The goal of a stand-up meeting is to get everybody in the team up to date on the current status including information about problems and solutions. Each communication activity has different needs for communication media that facilitates it. In summary, the creation of a communication strategy consists of three sub-activities:

*a) Plan communication activities:* The first step is to decide on the communication activities that need to be supported. Two types of activities can be distinguished: (1) Activities that should be carried out regularly (scheduled, e.g. daily stand-ups) and (2) activities that should occur on certain events (ad-hoc, e.g. coordination when question arises). Communication activities can be derived from the development process or from past experiences in similar projects. For each communication activity the information needs and exchange goals should be made clear.

*b) Plan media usage:* The next step is to decide on the communication medium that should be used for each communication activity. Choosing an adequate communication medium is a problem of its own. We can only refer to related work here, e.g. media richness theory [17] or media synchronicity theory [18]. Unfortunately, none of these theories incorporates all factors that we think have an influence on the media choice in a distributed setting. The following non-exhaustive list can be used as a reminder on what factors should be taken into account when choosing an adequate communication medium:

- Goal of the communication activity
- Content of information to be exchanged (e.g. code, requirements, decisions, …)
- Distribution
- Budget and time constraints
- Initial and regular setup times of medium
- Number of participants
- Differences in individual goals
- Differences in individual knowledge

For the sake of simplicity we only use media richness theory [17] here, because it provides an easy and intuitive way of evaluating and comparing communication media. Media richness theory states that communication media can be aligned a continuum of information richness. A decrease in richness will lead to a decrease in task outcome quality. In order of decreasing richness, examples of communication media are (1) face to face, (2) telephone, (3) personal documents (e.g. e-mail), or (4) impersonal documents (e.g. specifications). Our approach is to choose the richest medium available under distribution, budget, technological, and setup time constraints. One extension to the media richness theory in our approach is to add a specific channel for certain types of information content if feasible. For example, for some communication activities it can be useful to add a dedicated channel for desktop sharing to the otherwise plain audio based medium. For more examples see TABLE II.

*c) Create activity-specific FLOW Maps:* The final step in the creation of a communication strategy is to create FLOW Maps for each communication activity. These activity-specific FLOW Maps help to visualize all participating sites, how the information should be flowing between the sites, and the media to be used for cross-site communication. Examples of activity-specific FLOW Maps are given in Figure 3. and Figure 4. The creation of a FLOW Map is described in the following paragraph.

*3) Create overall target FLOW Map:* The final part of planning phase is the creation of an overall target FLOW Map. An overall target FLOW Map visualizes the project setup as planned with important information flows independent of specific communication activities. This way important people and documents can be identified early on and communication responsibilities can clearly be stated.

Based on the established team and the communication strategy the target FLOW Map can be created by performing the following 6 steps:

1. Draw dedicated areas for each participating site.
2. Draw a fluid store for each team member.
3. Draw solid stores for all documents or data stores that are essential to exchange project information.
4. Sketch a fluid flow between two people if they are supposed to exchange information on a regular basis. Add a label for the content of a flow. The communication strategy helps to identify regular flows and information needs. Put an arrow in if information will mainly flow in one direction. Sketch a fluid flow between a person and a document if the person is supposed to regularly solidify information in that document. Use different line widths to emphasize frequent or intense flows. The richness of the communication medium also influences the line width.
5. Sketch a solid flow between a document and a person if the document is supposed to be read regularly during the project.
6. Finally, the FLOW Map will be complemented by yellow pages information for each participant. In order to increase team awareness yellow pages information should contain, among others, name, picture, contact information, role in project, skills relevant for project, and experience with certain technologies.

The FLOW Map can now be used in further planning iterations to communicate and improve the plan and during the project to manage communication.

*C. Managing Communication*

In this Section we describe how to manage communication in distributed projects using the FLOW Map. Activities for *managing* communication in GSE are:

*1) Conformance Analysis:* The first step in managing communication is to determine whether the developers are complying with the communication strategy or not. In order to find out if the communication strategy is being followed as expected the communication channels need to be monitored. To do so, a process conformance template as presented in [19] can be used to further define the expected behavior and find indicators and metrics to identify violations against it. A conformance template adapted for communication strategy compliance is presented in TABLE I. Examples of filled in conformance templates are presented in TABLE III.

*2) Update FLOW Map:* A FLOW Map needs to be updated in order to visualize the current communication situation. A current FLOW Map can be compared to the target FLOW Map from the planning phase to detect deviations from the plan. Support for updating the FLOW Map is a tool issue as opposed to a conceptual issue. Hence, we do not elaborate on this here.

*3) Coordinate communication:* The final step in management is to use the violations found in conformance analysis and deviations from the target FLOW Map as a starting point to improve project communication. Finding a valid improvement is typically very specific to the given project situation. This area needs further research. Thus, we focus on the conformance analysis during management in this paper.

TABLE I. COMMUNICATION STRATEGY CONFORMANCE TEMPLATE

| Template Element | Description |
|---|---|
| Communication Activity | Name of the communication activity for which conformance analysis is to be performed. |
| Goal | Information exchange goal of the communication activity. This is important so that severity of violations can better be assessed. |
| Activity Definition | Description of the communication activity. Includes schedule or events from the communication strategy that trigger the activity. Description should be detailed enough to derive violations. |
| Collected Data | Data collected implicitly, manually, and automatically that can be used to derive violations. |
| Process Violations | Indicators and metrics that can be used to identify violations against the communication strategy. |

## IV. EVALUATION

We conducted a case study to evaluate our concept. We first present the evaluation goals and method. We then describe the project of the case study and its execution. Finally we show analysis results and discuss implications.

### A. Evaluation Goals & Method

In this paper we present the FLOW Map as a cognitive tool to plan and manage communication in GSE projects. To evaluate its usefulness we setup a case study to answer the following research questions:

*1) Feasibility for planning:* Can the FLOW Mapping approach be used to create a communication strategy and a FLOW Map and thus is it feasible to plan communication in a distributed project? To answer this question we used the FLOW Mapping approach to plan communication for the project of the case study presented here.

*2) Feasibility for management:* Can the FLOW Mapping approach be used to monitor communication and update the FLOW Map and thus is it feasible to manage communication in a distributed project? To answer this question we used conformance analysis [19] to monitor communication of the case study. Based on conformance analysis results we updated the FLOW Map during the project and provided active feedback to the study participants. Conformance analysis results will also partly answer question 1 since only a feasible communication plan can be followed correctly.

*3) Cost:* What are the initial and regular costs of the FLOW Mapping approach? Since cost considerations have a strong influence on the usefulness of a new approach we estimate the efforts needed during planning and management in the case study.

*4) Impact:* How valuable is the FLOW Mapping approach for project planning and management? Here we concentrate on qualitative data analysis. After the project we asked the project manager if the FLOW Mapping approach was helpful in planning communication. During the case study we asked the developers in a daily survey if they used the FLOW Map today and if they feel it was useful.

### B. Evaluation Setting

In this section we present the setting of the case study. We first describe the project's context and its development method. After that we discuss threats to validity.

*1) Distributed XP with students:* We apply the FLOW Mapping approach to a distributed agile programming project that is part of Leibniz Universität Hannover's (LUH) computer science curriculum. This course has been held in a non-distributed co-local setting before, with a strong emphasis on teaching the eXtreme Programming (XP) practices in a realistic environment [19, 20]. This time, we integrated students from Technische Universität Clausthal (TUC) in the course to create a distributed team. As described in [20], the core of the course is a block week that helps to implement the 40h-week-practice of XP and to create a realistic project environment. In preparation for this week, we conducted two 4h workshops to teach the fundamentals of XP. The students from TUC participated in the second workshop. Thus, both student groups got to know each other in person before start of the 40h XP week. We also had a 4h customer workshop to learn the requirements (distributed) and two more 4h slots for technical spikes (distributed). There were another four 4h slots after the XP week for finalization, post-mortem analysis, and discussion. The research presented in this paper focuses on the XP week, because we expected all XP practices to be in a steady state based on prior experience [19].

*2) Threats to Validity:* Every empirical study has to deal with validity threats. We conduct a single group study, so we cannot compare projects using FLOW Mapping to projects that use other approaches. The measures of the daily survey are qualitative so results might be subjectively biased. Another bias may result from evaluation apprehension since the team members knew they were being observed.

Regarding external validity there are threats according the selection of the study participants. We included 4 undergraduate and 4 graduate computer science (CS) students distributed over two locations. Both locations were in the same time zone and in the same country. The team was culturally diverse, but all students knew the German culture from living in Germany for at least a few months. 3 out of 8 students were non-native speakers of German. The two sub-teams had different educations from different CS schools and were at different levels in their CS curriculum. The students of the case study participated in the class mainly because they wanted to learn XP. Hence, there is a threat that these students were more communicative than the average CS student. Also, since the study was conducted in an agile setting we can only claim validity for agile environments. In summary, our results will most likely be valid for small, proximally distributed, culturally homogenous, agile teams.

### C. Evaluation Execution

In this section we present the execution of the case study. We start with a description of how we used FLOW Mapping to plan the project and finish with a description of the conformance templates we used and how we collected the data for conformance analysis.

*1) Planning the Communication:* Here we describe how we planned communication for the project of the case study using the steps presented in Section III.B.

*a) Establishing the team:* First, we had to identify and agree on the participating sites. We decided that students from LUH and TUC should be part of the team. We tried to establish a team of size 8 to 10, because we experienced this to be a good size for co-located XP teams in previous projects [20]. We also tried to balance sub-team sizes between sites. In the end the team consisted of 8 students, 4 located at LUH and 4 located at TUC. Since the LUH students already were in their graduate studies they had more advanced programming skills than the undergraduates from TUC. The common language for the project was German.

*b) Creating the communication strategy:* In a brainstorming session we derived a set of communication activities that would be needed for cross-site communication during the project (see TABLE II. left column). Since XP was used for development we planned for daily stand-up meetings to coordinate development, planning games for iteration plan-ning, and acceptance tests for iteration validation. In addition to the activities that can directly be derived from XP we planned for daily wrap-up meetings that should mainly be used for educational wrap-ups, acceptance tests to validate each User Story, ad-hoc informal coordinative meetings, ad-hoc informal collaborative meetings, and finally status updates to broadcast changes of work items and pair compositions. The last three communication activities were specifically incorporated in the strategy since in distributed settings these types of communication do not happen as naturally as in the co-located setting. We distinguish between informal coordination and collaboration because coordinative information exchange usually has less requirements for the communication medium than creative collaborative work.

As stated in the concepts section we used a slight modification of the media richness theory [17] for planning media usage for simplicity reasons. Our approach was to choose the richest medium available under distribution, budget, technological, and setup time constraints. If appropriate we added a dedicated channel for specific contents. The richest medium that was available at both sides and could be used without extra monetary costs was high quality (HQ) video conferencing. Hence, we used this HQ video system for meetings involving the whole team, namely stand-up, wrap-up, planning game, and acceptance test of the iteration. Since prioritization of User Stories is part of the planning game we added a dedicated channel to share User Stories between sites during the planning game. We did not find a dedicated tool for distributed User Story management that fulfilled our needs. Instead, we used a shared mind map. We also included a shared desktop to the meetings for acceptance testing the iteration, because the on-site customer needed to be able to interact with the product. We planned to use Skype for cross-site communication that takes place during development. Again, the medium for the acceptance test was extended by a shared desktop. A shared desktop is also helpful when it comes to creatively collaborating on something, e.g. source code or architecture. For the informal communications we did not prescribe whether Skype calls or Skype text chats should be used. Since Skype was used at the developer workstations the Skype status message functionality was available for broadcasting status changes.

TABLE II. COMMUNICATION STRATEGY OF CASE STUDY

| Communication activity | Schedule / event | Communication media | FLOW Map |
|---|---|---|---|
| *Stand-up[a]* / *Wrap-up[a]* | Every morning / evening | HQ video conference | - |
| *Planning game[a]* | Start of iteration (~ every 2. day) | HQ video conference with shared mind map | Figure 3. |
| *Acceptance test of iteration* | Iteration completed | HQ video conference with shared desktop | - |
| *Acceptance test of user stories[a]* | User story completed | Skype call with shared desktop | - |
| *Informal collaboration* | Ad-hoc | Skype call/chat and desktop sharing | Figure 4. |
| *Informal coordination* | Ad-hoc | Skype call / chat | - |
| *Status update[a]* | Status change | Skype status | - |

a. Conformance analyses of marked activities are presented in this paper

The last step in the creation of a communication strategy is to create activity-specific FLOW Maps. These maps visualize sites, participants, important documents, media for cross-site communication and information flows of a specific communication activity. Thus, these FLOW Maps help to identify shortcomings in the plan. Also, they can be used to communicate the plan to the developers at start of the project. Figure 3. shows the FLOW Map for the planning game activity. A dedicated moderator is steering the meeting. He also acts as a scribe when documenting the requirements as user stories in the shared mind map tool. When written down, on-site customer and developers can read the documented user stories. The developers estimate efforts needed for each user story. Based on these estimates the customer can prioritize or change requirements.

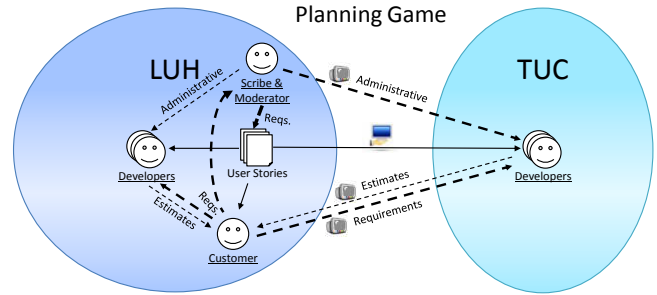

Figure 3. FLOW Map of a planning game

A second activity-specific FLOW Map is presented in Figure 4. The pairs communicate via audio or instant messaging. If they collaborate on a certain work item, e.g. source code or a UML diagram, they can modify the remote work item via desktop sharing.

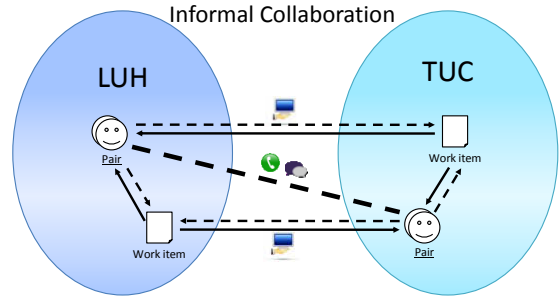

Figure 4. FLOW Map of an informal collaborative meeting

The planned communication activities, their schedule or trigger, selected FLOW Maps, and communication media of the case study are summarized in the communication strategy in TABLE II.

*a) Creating the overall target FLOW Map:* The final step of the planning phase is the creation of the overall target FLOW Map including yellow pages information of all participants. Figure 1 depicts the overall target FLOW Map as used in our case study. Both sites have 4 developers and a dedicated coordinator. The on-site customer is located at LUH. LUH is responsible for the ticket management (Trac) and version control (SVN) systems. We consider local communication between pairs to be strong. We planned that most information across sites will flow directly between pairs since we wanted to stay as agile as possible in the distributed setting. Another communication medium across sites is the source code itself via the version control system SVN. The coordinator at LUH also was a technical and an XP specific coach. Yellow pages information shows names, pictures and contact information for everyone. In addition, the developers are mapped to a pair and to their current work item (User Story IDs and names).

*2) Managing Communication:* Knowing how good the development team is communicating is fundamental for managing communication. Only if a project manager knows what works well and what does not she can make the right decisions. A first indicator for communication quality is whether the developers are communicating according to the plan, i.e. the communication strategy. We use conformance analysis as presented in [19] to measure compliance to the communication strategy. In TABLE III. three examples of filled in conformance templates (see TABLE I. ) as used in the case study are presented.

### D. Data Collection and Descriptive Statistics

In this section we give an overview of the data we collected during evaluation. We monitored and collected communication activities as listed in the communication strategy based on media usage of the team. When possible we used log files of the tools facilitating a communication to derive type, start time, end time, and participants of an activity. We could use logs from Skype and the video conferencing system. Since the Skype logs do not contain status message changes we had to develop our own tool to collect this data. Version control system events could be derived from SVN history data. The developers had to fill in a template containing pair names and User Story ID for each commit. The rest of the collected data, like communications with the customer, pair switches, and User Story assignments, was collected manually by dedicated observers. Unfortunately, we did not get access to the desktop sharing log files, since the tool we used for desktop sharing was hosted externally. Figure 5. summarizes the frequency of communication media usage during the XP week independent of communication durations and richness of used medium. Figure 6. summarizes the durations of communication media usage on average for each developer per day. We did not include events that do not have a duration or for which it is very difficult to determine the duration like Skype text chats or SVN commits in this figure.

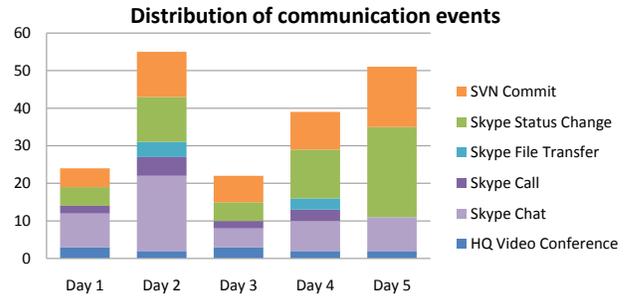

Figure 5.   Frequency of media usage during XP week

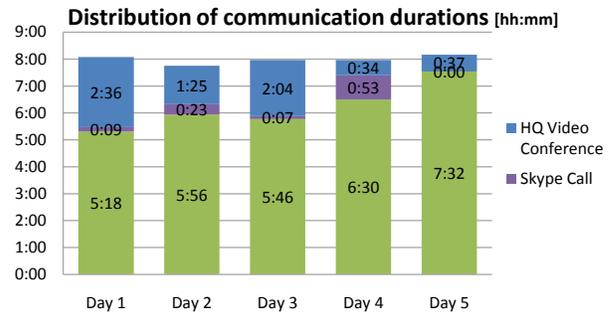

Figure 6.   Durations of communication per developer during XP week

TABLE III.   SELECTED CONFORMANCE TEMPLATES OF CASE STUDY

| Communication Activity | Status update | Acceptance test of User Stories | Scheduled activities: Stand-up, wrap-up, planning game |
|---|---|---|---|
| Goal | Increase awareness on who is working with whom on what task | Validate development outcome with customer needs | Each activity has specific goals. Overall: Distribute important information to all team members regularly |
| Definition | Developers should use Skype status messages to broadcast who is working with whom on which User Story in a timely manner. The status message should contain User Story ID and the names of the pair programmers. | Developers present their work to the On-Site Customer if they think they completed a User Story. The customer decides, if the User Story is completed. | Overall: Scheduled communication activities should be carried out regularly according to the schedule, i.e. every morning (stand-up), every evening (wrap-up), and at start of iteration (planning game) respectively. |
| Collected Data | Skype status log for each workstation containing: timestamp and status message and status change events (pair switches, assignment of new User Stories) | Communication events with customer, SVN commits with User Story ID | Start and end times of scheduled communication activities. |
| Violations | **Temporal:** (1) Status message not updated for more than one hour (2) Status message suggests that a developer is working in two pairs concurrently **Qualitative:** (1) Incomplete information, e.g. User Story ID missing. | **Temporal:** (1) User story finished before acceptance test (i.e. no communication event between pair and customer before commit) **Qualitative:** (1) User story finished without acceptance test (i.e. no communication event between pair and customer at all) | **Temporal:** (1) Communication activity is being carried out too early (previous day) or too late (next day). **Qualitative:** (1) Communication activity is not being carried out at all. |

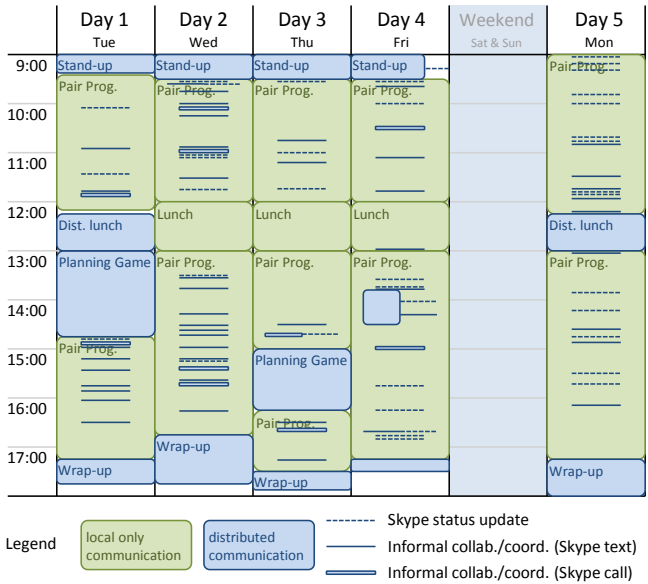

Figure 7. Overview of communication during XP week

Based on the media usage data we created an overview of project communication as shown in Figure 7. Local only communication is depicted in green and cross-site communication is depicted in blue. Labels indicate names of communication activities. Unlabeled activities can be distinguished by the shape of the horizontal line or box. Dashed lines show Skype status updates. Solid lines represent informal communication via Skype text chats. Blue boxes show informal communication via Skype calls.

*1) Communication Strategy Conformance Results:* In this section we present conformance analysis results for the three communication activities listed in TABLE III.

Conformance to the communication activity "status update" was monitored by logging Skype status messages of the participating development groups on LUH and TUC site. Violations against the practice are situations where developers either forget to maintain their Skype status message, or if the information is incomplete. The first type of violation (temporal) includes situations where developers do not have or do not update a status message for more than one hour. Situations where developers forget to update their names in the status message (and therefore appear to be working in two teams at the same time), are considered temporal violations. The second type of violation (qualitative) includes scenarios where developers post a Skype status, but the status is incomplete, e.g. does not contain a User Story number and/or the names of the developers. Figure 8. shows the temporal and qualitative violations for each of the five development days. From day 1 to day 2 relative compliance slightly increased, reached 100 % on day 3, and then decreased on day 4 and 5. On average the compliance level for the activity "status update" was 79 % with 8 % temporal violations and 13 % qualitative violations.

Conformance to the communication activity "Acceptance Test of User Stories" was measured by manually collecting communication events between on-site customer and developers. In an acceptance test the on-site customer examines the features that were developed for a certain User Story and validates these features against the customer's real needs. Thus, an acceptance test can never be considered successful without any communication between developers and customer. Violations against this practice are situations where developers communicate with the customer after completion of a User Story or where developers do not communicate with the customer about a User Story at all. We used SVN commit messages to identify completed User Stories and compared these results with communication events with the customer. Out of 16 SVN commits that were marked as "User Story completed" two violated the conformance criteria. One commit was performed for a User Story without any communication between developers and customer at all (qualitative) and one commit was performed before communication with the customer took place (temporal). Therefore the overall compliance level for the activity "acceptance test of User Story" was 88 % with 6 % temporal violations and 6 % qualitative violations.

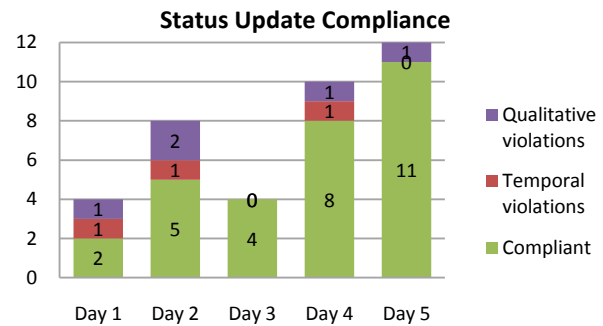

Figure 8. Conformance analysis of activity "status update"

Conformance analysis for the scheduled activities was performed by monitoring their start and end times. Out of the 13 scheduled communication activities – 5 stand-ups, 5 wrap-ups, and 4 planning games – we found 2 (15 %) violations. At the last development day there neither was a stand-up meeting nor a planning game although both were on schedule for that day. So the compliance level to the communication schedule was 85 %.

*2) Survey Results:* First, we asked the project manager who was responsible for planning and managing communication for the project whether the FLOW Mapping approach was useful for (1) planning and (2) managing communication. Following are his key points about using FLOW Mapping for planning:

- Seeing participants at different sites helped to think about and talk about role assignments.
- Seeing desired information flows helped to schedule and prioritize communication activities.

The project manager's key points about management:

- Seeing assigned tasks and pairs was beneficial.
- Seeing images along with names of the developers enables me to call everybody by name from day one.

To answer the second part of research question 4 we conducted a survey. At the end of each day the developers had to fill in a questionnaire that included questions about whether they used the FLOW Map that day and how useful they think it was. Figure 9. shows that the FLOW Map was used by 75 % of the developers in the beginning of the project, then the usage rate declined to 38 % on day 4, and it rose again to 50 % after the weekend on day 5. Figure 10. shows how useful the developers perceived the FLOW Map. Perceived usefulness declined monotonic from 100 % on day 1 to 38 % on day 5. To correctly interpret these results it is important to know that the FLOW Map did not show current User Stories for days 4 and 5 because we could not keep up with updating the map 4 due to too frequent changes.

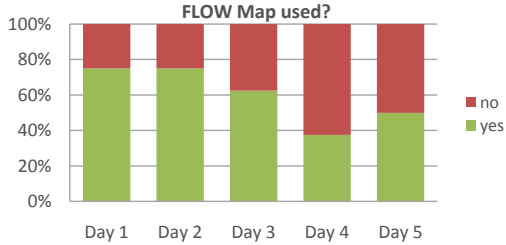

Figure 9. Usage of FLOW Map

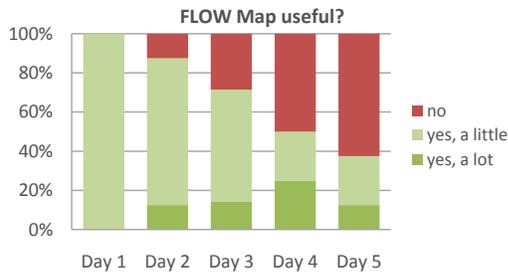

Figure 10. Perceived usefulness of FLOW Map

### E. Discussion

In this section we discuss the results of the case study in correspondence with the research questions from Section IV.A. We discuss the results in reverse order because some results of the latter questions help to answer the former.

*1) Impact:* Results of the survey suggest that the FLOW Map was used and perceived to be useful during the project especially in the beginning of the XP week. Hence, information about distributed team members seems to be valuable at project start when distributed teams still grow together and awareness has to be developed both for developers and for project management. Also, only an up to date map can make an impact. Towards the end of the XP week when pairs and user Story assignments switched more frequently our manual update process could not keep up anymore. The FLOW Map became out of date. To prevent this in future projects we started to develop a web based tool that visualizes the FLOW Map and automatically and semi-automatically monitors communication activities to keep the visualization up to date.

*2) Cost:* A new approach is only useful if its benefits outweigh its costs. Here we consider cost in terms of time needed to plan and monitor communication using FLOW Mapping. TABLE IV. gives an overview of the times needed for planning the project, for preparing and executing conformance analysis, and for keeping the FLOW Map up to date during the project. In summary, when using the FLOW Mapping approach communication of a distributed project can be planned in one workday. Preparation of conformance analysis can also be accomplished in one or two work days, if no new monitoring tools need to be developed. Execution of conformance analysis on the other hand can become very expensive if manual data collection is needed. For larger and longer projects it might pay off to initially invest in monitoring tools and save regular monitoring costs during the project. Depending on the number of monitored communication events analysis of the collected data can also be costly. In our case we could off-shore analysis of conformance data since there was a 6 h time zone difference between the development sites in Germany and the conformance analysis expert in the USA. This way conformance analysis results for one day were readily available at start of the next day.

TABLE IV. TIMES NEEDED FOR FLOW MAPPING

| Process step | Time estimates | Freq. |
|---|---|---|
| Plan: establish team | 4h coordinating calls and mails | 1 |
| Plan: comm. strategy | 2h brainstorming meeting | 1 |
| Plan: overall FLOW Map | 15 min. sketch, 1 h polished | 1 |
| Conformance analysis → fill in templates | 30 min. each | # comm. activities |
| Conformance analysis → prepare data collection | 1 d develop Skype status monitor, 1 h setup and connect Skype accounts, 2 h data collection forms, 15 min. SVN commit template | # data sources |
| Conformance analysis → collect data | 8 h manual data collection, e.g. communication w/ customer | daily |
| Conformance analysis → analyze data | 1 h each template | daily |
| Update FLOW Map | 10 min. each status change | daily |

*3) Feasibility for management:* The results of conformance analysis show that it is feasible to monitor pre-defined aspects of communication when running a distributed project. The results were used to actively steer communication behavior or change the plan, depending on what is more appropriate in a given situation. For example, detecting a violation against the communication schedule on the last day of development and not enforcing the schedule but allowing developers to cleanly finish their work (i.e. changing the plan) might be a good idea in this situation. The results about the usefulness of the FLOW Map (Figure 10. ) suggest that it can be helpful for the developers but only if it is current. Towards the end of the XP week we could not update the map fast enough due to the frequent changes and our slow manual update process. Hence, tool support is needed for fast paced development projects. Nowadays a lot of project information is already being maintained electronically, e.g. issue tracking systems for task management. Therefore, connecting these data sources to the visualization of a FLOW Map should be feasible to implement in a future tool. E.g., our Skype status message monitor could be extended to send the names of a pair as noted in the status message to a visualization tool.

*4) Feasibility for planning:* By using FLOW Mapping to plan communication for the case study we could show that the concept is feasible. Furthermore, actual cross-site communication and high compliance levels between 79 % and 88 % show that the plan itself is feasible, too.

## V. Conclusion & Outlook

Informal communication is an important part of software projects. In distributed settings informal communication is hindered. Therefore, incorporating informal communication during planning and management can help to improve distributed projects. We presented the FLOW Mapping approach and FLOW Maps as the process and cognitive tool that guide planning and management of communication in distributed projects. Since FLOW Mapping stems from FLOW research it incorporates informal communication at all stages.

In a case study we showed that it is feasible to plan and manage communication in distributed projects using the FLOW Mapping approach. The steps for planning helped to establish a distributed team, to create a communication strategy, and to create an overall target FLOW Map. During the project conformance analysis showed that the strategy was being followed with compliance levels between 79 % and 88 %. We also discussed costs in terms of time needed for planning and management. Being able to plan communication in one work day is acceptable. Execution of conformance analysis can be expensive if data about communication behavior is needed that cannot be monitored automatically. Using the logs of the communication media that were used for cross-site communication saved a lot of manual effort.

Another application of the FLOW Map is the improvement of communication during a project from the developers' point of view, i.e. by increasing team awareness. Our results showed that developers find the FLOW Map useful especially in the early phases of a distributed project, when team building takes place.

In future research efficiency of conformance analysis needs to be improved in order to further reduce human effort. To reach this goal new information flow metrics are needed. For example a metric about whether information in a document really is solid, i.e. all team members can understand it, would help to monitor documentation quality.

## Acknowledgment

We would like to thank our project partners at Technische Universität Clausthal, the participating students, and everybody who helped in the process of planning and executing the case study.

This work was partly funded by the German Research Foundation (DFG project InfoFLOW, 2008-2011) and the Ministry of Science and Culture of Lower Saxony (project GloSE, 2010-2013).